\documentstyle[twocolumn,prb,aps]{revtex}
\begin{document}
\title{Phase Coexistence Properties of Polarizable 
Stockmayer Fluids}
\author{Kenji Kiyohara, Keith E. Gubbins 
and Athanassios Z. Panagiotopoulos\cite{azp}}
\address{School of Chemical Engineering,\\
   Cornell University, Ithaca, NY 14853-5201, USA}
\date{\today}
\maketitle

\section{Abstract}

	We report the phase coexistence properties of polarizable 
Stockmayer fluids of reduced permanent dipoles 
$|\bbox{m}_0^*|=$ 1.0 and 2.0 
and reduced polarizabilities $\alpha^*=$ 0.00, 0.03, and 0.06,
calculated by a series of grand canonical Monte Carlo simulations with
the histogram reweighting method.  
In the histogram reweighting method, the distributions  
of density and energy calculated in Grand Canonical Monte Carlo simulations 
are stored in histograms and analyzed to construct  
the grand canonical partition function of the system.  
All thermodynamic properties are calculated from the 
grand partition function.  
The results are compared with 
Wertheim's renormalization perturbation theory.  
Deviations between theory and simulation results 
for the coexistence envelope are near 2\% for the lower dipole 
moment and 10 \% for the higher dipole moment we studied.

\section{Introduction}

	Stockmayer fluids \cite{stm} have long been studied as models 
for fluids with permanent dipoles, such as water, 
ammonia, or methyl chloride. Thermodynamic 
properties for these fluids have been calculated by theory and simulations.  
Attempts have been made to model real dipolar fluids by fitting 
potential parameters of Stockmayer fluids.
Rowlinson \cite{jsr} fitted experimental 
second virial coefficients to theoretically calculated
ones to find Stockmayer potential parameters $\epsilon$, $\sigma$, 
and $|\bbox{m}|$ for some dipolar fluids.   
Van Leeuwen \cite{vL} fitted 
experimental coexistence curves to results from 
computer simulations.   
Agreement between the Stockmayer potential parameters 
calculated from second virial coefficients and phase coexistence
data is only qualitative.  
One of the reasons why the agreement is not quantitative is that 
fitting of the second virial coefficient gives the parameters 
for the fluid at the limit of zero density while 
fitting of the coexistence curve gives the parameters for 
the dense fluid.  The interaction 
of dipolar molecules can significantly change depending on the 
density or temperature due to the redistribution of electron 
density within a molecule in response to changes in the molecular 
environment (electrostatic induction effect).
It is essential to account for this effect when 
phase coexistence properties of dipolar fluids are calculated 
because electrostatic induction is much stronger in the liquid 
phase than in the gas phase, and molecular interactions 
cannot be accurately modeled by the same 
(non-polarizable) model and parameters 
for both phases.  
One way to include the electrostatic induction effect 
on a model of polar fluids is to introduce polarizability.  

Wertheim \cite{wert1} has studied the effect of polarizability on 
thermodynamic properties by using a graph-theoretical approach.  
His renormalization perturbation theory \cite{wert2} was then extended 
to mixtures by Venkatasubramanian et al. \cite{gubbins1}.
Patey et al. \cite{patey} performed Monte Carlo simulations to test 
the free energy calculation by Wertheim's theory for 
hard spheres with moderately large reduced dipoles 
$ | \bbox{m}_0^*| = |\bbox{m}_0|/ \sqrt{kT d^3} =$ 1.0 
(where $\bbox{m_0}$ is the permanent dipole moment)
and reduced mean polarizabilities 
of $\alpha ^* =\alpha / d^3 =0.03, 0.06, 0.1$  
(where $k$ is Boltzmann constant, 
$T$ is the absolute temperature, 
$d$ is the hard sphere diameter and
$\alpha$ is polarizability).
Venkatasubramanian et al. compared 
theoretically calculated coexistence properties 
with experiment \cite{gubbins1}
for Stockmayer fluids with reduced dipoles of 
$|\bbox{m}_0^*|=|\bbox{m}_0| / \sqrt{\epsilon \sigma^3} \simeq 1.0$ 
and reduced polarizabilities 
of $\alpha ^* = \alpha / \sigma^3 \simeq 0.06$  
($\epsilon$ and $\sigma$ are the Lennard-Jones parameters).
The results of these studies 
agree reasonably well with Wertheim's theory.  
Vesely \cite{vesely} performed molecular dynamics simulations of polarizable 
Stockmayer fluids and calculated the effect of polarizability 
on thermodynamic properties such as internal energy or pressure.  
Smit et al. \cite{smit1} and van Leeuwen et al. \cite{smit2}
performed Gibbs ensemble Monte Carlo simulations of
coexistence properties for non-polarizable Stockmayer fluids.  However, 
simulation studies of the coexistence properties of 
polarizable Stockmayer fluids 
have not been previously published to our knowledge.  

In this paper, we present results from Grand Canonical Monte Carlo 
(GCMC) simulations of 
Stockmayer fluids with and without polarizability for the vapor-
liquid phase coexistence properties.  
The results are compared to the renormalization perturbation theory 
by Wertheim \cite{wert2,gubbins1}.  
We study the models with  reduced dipoles of 
$|\bbox{m}_0^*|=$ 1.0 and 2.0
and reduced polarizabilities of $\alpha ^*=$0.00, 0.03, and 0.06.  
Examples of estimates of the parameters for real fluids are 
$|\bbox{m}_0^*|=$ 1.84 and $\alpha ^*=$0.08 for water 
\cite{jsr} and 
$|\bbox{m}_0^*|=$ 1.03 and $\alpha ^*=$0.06 for methyl chloride 
\cite{gubbins1}.   
Since applications of the histogram reweighting method to
phase coexistence of fluids have only recently
started appearing\cite{wilding}, we begin this paper by 
discussing the principle of the method and issues related to its 
practical application to predict phase coexistence curves 
at temperatures significantly below the critical point.  

Deviations between theory and simulation results 
for the coexistence envelope are near 2\% for the lower dipole 
moment and 10 \% for the higher dipole moment studied.

\section{GCMC Histogram reweighting method}

Determination of phase coexistence by Monte Carlo simulation requires 
either implicit or explicit calculation 
of the free energy of a system.  The Gibbs 
ensemble method \cite{panagiot} is used for determination of phase 
equilibrium by 
implicitly minimizing the total free energy of the system, which is 
separated into two phases.  
Non-Boltzmann sampling methods (such as thermodynamic scaling)
\cite{valleau1,valleau2} and the test  
particle insertion method \cite{widom} are among the methods 
for explicit calculation of free energy of the system. 
Ferrenberg and Swendsen \cite{fs}
proposed the use of the distribution of energy and density calculated 
in grand canonical Monte Carlo (GCMC) simulations. 
We refer to this method as ``the histogram reweighting method.''  
The method has rarely been applied to continuous-space fluids, 
with the exception of a recent study by Wilding for the Lennard-Jones 
fluid \cite{wilding}.  We chose the histogram reweighting method for
this study because we found it to be computationally more efficient
than the other available methods for our systems.

One of the attractive features of the histogram reweighting method is
that it can be used to construct the grand canonical partition function, which in
turn can be used to obtain all thermodynamic properties, including the free energy or 
coexistence properties.  Moreover, a single simulation at a given chemical potential $\mu$ and 
temperature $T$ can give the thermodynamic properties at 
a range of $\mu'$ and $T'$,  by virtue of the scaling properties in 
the variables (see section on "Theoretical Basis" below).
In calculating coexistence properties, it is not necessary to observe 
phase coexistence in a single GCMC simulation, 
since they are calculated by analyzing the grand canonical partition 
function constructed by combining the histograms.  
GCMC simulation is appropriate for calculating thermodynamic properties 
for a range of densities for molecular fluids, because it does not
require computationally expensive volume changes.  The polarizable 
Stockmayer potential does not scale simply with volume.  

\subsection{Theoretical Basis}

	The basic concept behind the histogram reweighting method of 
Ferrenberg and Swendsen \cite{fs} for GCMC simulations 
is reviewed here.  
The grand canonical 
partition function of a system with chemical potential $\mu$, volume $V$, 
and temperature $T$ is written as 

\begin{eqnarray}
	\Xi(\mu,V,T)  = \sum_N \sum_{U_N} exp[(N \beta \mu)-\beta U_N] \Omega (N,V,U_N)
\end{eqnarray}

\noindent where $\Omega (N,V,U_N)$ is the number of microstates for
number of particles $N$, volume $V$ and energy $U_N$;
the notation $U_N$ emphasizes that the energy level depends 
on the number of particles.  
We define the chemical potential $\mu$ by 
\begin{equation}
	\beta \mu = ln \ z 
\end{equation}
where $z$ is the activity.  
$\beta$ is the inverse temperature ($\beta =1/kT$) 
, where $k$ is Boltzmann's constant.  
The $\displaystyle \sum_N$ denotes the summation over $N$ from 0 to 
infinity, and $\displaystyle \sum_{U_N}$  denotes the summation over all energy levels for 
each $N$.   
We perform GCMC simulations by Norman and Filinov's method \cite{nf} and 
store the number of observations 
of particular $N$ and $U_N$ in a two dimensional histogram $f_{\mu,V,T}(N,U_N)$, 
which is related to the components of the grand canonical partition function, 
with a simulation-specific constant $C$, by  
\begin{equation}
	f_{\mu,V,T}(N,U_N) \cdot C = exp[(N \beta \mu)-\beta U_N] \Omega (N,V,U_N)
\end{equation}

\noindent The thermodynamic average of a property $X$ is calculated by 
\begin{equation}
	<X>_{\mu,V,T}   = \frac{\displaystyle\sum_N \sum_{U_N} X(N,U_N) f_{\mu,V,T}(N,U_N)}
	{\displaystyle\sum_N \sum_{U_N} f_{\mu,V,T}(N,U_N)}.
\end{equation}

Next, let us consider the grand canonical partition function for a different thermodynamic state 
with chemical potential $\mu'$ and temperature $T'$, which is written as 

\begin{eqnarray} \label{pf}
\Xi(\mu',V,T') = \nonumber \\
= \sum_N \sum_{U_N} { exp[(N \beta' \mu')-\beta' U_N] \Omega (N,V,U_N) }  \nonumber \\
\nonumber \\
= \sum_N \sum_{U_N} exp[N(\beta' \mu'-\beta \mu) -(\beta'-\beta) U_N] \cdot  \nonumber \\
\cdot exp[(N \beta \mu)-\beta U_N]\Omega (N,V,U_N)   \nonumber \\ 
\nonumber \\
 =  \sum_N \sum_{U_N} exp[N(\beta' \mu'-\beta \mu)-(\beta'-\beta) U_N] \cdot  \nonumber \\
   \cdot f_{\mu,V,T}(N,U_N) \cdot C 
\end{eqnarray}

\noindent The thermodynamic average of a property $X$ is then 

\begin{eqnarray} \label{x}
<X>_{\mu',V,T'}   =  \nonumber \\ 
\nonumber \\
= \frac{ \displaystyle\sum_N \sum_{U_N} X(N,U_N) 
 W f_{\mu,V,T}(N,U_N)}
{\displaystyle \sum_N \sum_{U_N} W f_{\mu,V,T}(N,U_N)}
\end{eqnarray}

\noindent where $W = exp[N(\beta' \mu'-\beta \mu)-(\beta'-\beta) U_N]$.
As can be seen from these equations, once we know the components of 
the grand canonical partition function $f_{\mu,V,T}(N,U_N)$ at a 
thermodynamic state ($\mu$, $V$, $T$), 
we can construct a grand canonical partition function at 
a different thermodynamic state 
($\mu'$, $V$, $T'$) by reweighting each component with 
$exp[N(\beta' \mu'-\beta \mu)-(\beta'-\beta) U_N]$.  
Then, we can calculate the thermodynamic properties 
at the new state ($\mu'$, $V$, $T'$) by averaging a thermodynamic variable with 
the reweighted components of the grand canonical partition function.  
Therefore, once we have a two dimensional histogram $f_{\mu,V,T}(N,U_N)$ 
from a GCMC simulation at a state ($\mu$, $V$, $T$), 
any thermodynamic property at any state ($\mu'$, $V$, $T'$). 
The constant $C$, which is unknown, but can be determined 
as described below, is unimportant in averaging 
a property $X$ as in equation \ref{x} because it cancels out in 
the numerator and the denominator.

\subsection{Determination of phase coexistence}

	In the grand canonical ensemble at a sub-critical temperature, the 
vapor-liquid coexistence can be determined by finding the chemical potential 
that gives the same pressure for both phases.  We define the pressure $P$
at a thermodynamic state ($\mu$, $V$, $T$) by 

\begin{equation} \label{prs1}
	P_{\mu, V, T} = \frac{1}{V} ln \ \Xi(\mu, V, T) 
\end{equation}

\noindent If a histogram is made for a thermodynamic state ($\mu$, $V$, $T$), 
the pressure at ($\mu'$, $V$, $T'$) 
is calculated by

\begin{eqnarray} 
	P_{\mu', V, T'}= &&  \nonumber \\
= \frac{1}{V} ln \sum_N \sum_{U_N} 
exp[N(\beta' \mu'-\beta \mu)-(\beta'-\beta) U_N] \cdot \nonumber\\
\label{prs2} \cdot f_{\mu,V,T}(N,U_N) \cdot C 
\end{eqnarray}

\noindent where equations \ref{pf} and \ref{prs1} are used.  
At a subcritical temperature, the pressure of each phase is thus 
calculated except 
for the constant $C$.  
The chemical potential $\mu'$ that gives the same pressure for both phases 
is then found, and the phase coexistence thus determined.  
When it is necessary to calculate the absolute value of the partition function, 
the simulation specific constant $C$ needs to be determined by 
a different means.
In this study, the absolute value of the constant $C$ is fixed 
by calculating the pressure in the original 
simulation using the virial theorem 
\begin{equation}
	P =\frac{<N>_{\mu,V,T}}{V}kT + 
\Big< \frac{d U}{d V} \Big>_{\mu,V,T}
\end{equation}
and equating that with the pressure 
calculated by equation \ref{prs2}.  

\subsection{Combining Results of Several Simulations}

Calculation of the thermodynamic properties at vapor-liquid coexistence
requires the histogram over a wide range of $N$ and $U_N$.  
However, it is difficult to get $f_{\mu,V,T}(N,U_N)$ with 
good statistical accuracy over a wide range of $N$ and $U_N$ 
from one simulation.   In 
the method by Norman and Filinov \cite{nf} or in any Metropolis scheme 
\cite{metro}, the configurations 
relevant to the given thermodynamic state ($\mu$, $V$, $T$) are 
preferentially sampled and 
those relevant to other states are not sampled well.  
Therefore, we perform several simulations at different thermodynamic states and 
combine the information to construct the histogram $f_{\mu,V,T}(N,U_N)$, which 
is then given with good accuracy over a wide range of $N$ and $U_N$.  
In the initial attempts to sample a wide range of $N$ and $U_N$, we 
perform GCMC simulations at a temperature slightly above 
the critical point with various chemical potentials.  The details are 
described in section 5.   

Combining two simulation results is done by fixing 
the ratio of the simulation specific constants $C$ for two simulations at 
different thermodynamic states (denoted by subscripts 1 and 2) by imposing 

\begin{eqnarray}
 \sum_{U_N} exp[N(\beta \mu-\beta_1 \mu_1)-(\beta-\beta_1) U_N] \cdot \nonumber \\
\cdot f_{\mu_1,V,T_1}(N,U_N) \cdot C_1 
\nonumber\\
= \sum_{U_N} exp[N(\beta \mu-\beta_2 \mu_2)-(\beta-\beta_2) U_N] \cdot \nonumber \\
f_{\mu_2,V,T_2}(N,U_N) \cdot C_2 
\end{eqnarray}

\noindent at the same $N$ and $T$: if the original simulations are 
performed at different temperatures, one or both histograms need 
to be reweighted so that the two histograms are compared at the same 
temperature.  It is necessary to choose $N$ and $T$ 
for which the relevant configurations are sampled by both simulations.
For convenience, we choose $T$ as the average of the temperatures for 
the two original simulations and $N$ as the number of particles at 
which the density distributions at temperature $T$ overlap most.  
A more sophisticated way of combining multiple simulation 
results,  utilizing 
information for a range of $N$ instead of one $N$,
has been proposed by Ferrenberg and Swendsen \cite{fs2}.
The simpler method described above was found to be satisfactory for our systems.  

\subsection{Comparison to the Gibbs Ensemble method}
Both the GCMC histogram reweighting method and the Gibbs ensemble method 
\cite{panagiot} can 
be used to obtain the coexistence properties of a system such as the polarizable
Stockmayer fluid of the present study.  At first sight, it would seem that the
Gibbs ensemble method is simpler to implement, as it requires only a single 
simulation per temperature at which coexistence is to be determined, rather than
the series of simulations required by the GCMC histogram method.  To
determine the complete phase diagram at a series of temperatures both
methods require several simulations.  However, we have found that 
statistical uncertainties for the coexistence properties from the GCMC 
method seem to be significantly smaller than for a Gibbs-ensemble
determination of the phase behavior for comparable amounts of computational
time expenditure.  This statement is supported by the small statistical
uncertainties of the coexistence properties calculated in Tables III and IV, which
would have required prohibitively long Gibbs ensemble calculations.

\section{Model}

	A molecule of a Stockmayer fluid is a Lennard-Jones 
interaction site with an embedded point dipole at the center of the molecule.  
For polarizable models, polarizability is introduced also at the center of 
the molecule \cite{patey,vesely}.  Assuming that the induced 
dipole moment is linearly 
dependent on the local electric field at the center of the molecule, the 
total dipole moment of molecule $i$ is written as 

\begin{equation}
\bbox{m}_i=\bbox{m}_{0,i}+\bbox{\alpha} \cdot \bbox{E}_i, 
\end{equation}

\noindent where $\bbox{m}_i$ is the total dipole moment, $\bbox{m}_{0,i}$ 
is the permanent dipole moment, 
$\bbox{E}_i$ is the local electric field at the center of the molecule, 
and $\bbox{\alpha}$ is the polarizability tensor.  
The local electric field is calculated as \cite{gg}

\begin{equation}
\bbox{E}_i=\sum_{j \neq i} \bbox{T}_{ij} \cdot \bbox{m}_j , 
\end{equation}

\noindent where $\bbox{T}_{ij}$ is the dipole tensor for 
molecules $i$ and $j$.  If the vector 
connecting the centers of molecules $i$ and $j$ is written as 
$\bbox{r}_{ij}$ and 
the unit tensor as $\bbox{I}_{ij}$, the dipole tensor is defined as 

\begin{equation}
	\bbox{T}_{ij}=\frac{3 \bbox{r}_{ij} \bbox{r}_{ij}}{{r_{ij}}^5}
		-\frac{\bbox{I}_{ij}}{{r_{ij}}^3} 
\end{equation}
\noindent	where 
\begin{equation}
	r_{ij}=|\bbox{r}_{ij}| 
\end{equation}

\noindent The total interaction of a Stockmayer fluid is $U=U_{LJ}+U_{dp}$ , 
where

\begin{equation}
	U_{LJ} = \sum_{i\neq j} 4 \epsilon 
	\Big[(\frac{\sigma}{r_{ij}})^{12}-(\frac{\sigma}{r_{ij}})^6\Big]
\end{equation}

\begin{eqnarray}
	U_{dp} = -\frac{1}{2}\sum_i \bbox{m}_i \cdot \bbox{E}_i 
+ \frac{1}{2} \sum_i \bbox{E}_i \cdot \bbox{\alpha} \cdot \bbox{E}_i \nonumber \\
= - \frac{1}{2}\sum_i \bbox{m}_{0,i} \cdot \bbox{E}_i 
\end{eqnarray}

In these equations, $U_{LJ}$ and $U_{dp}$ denote the Lennard-Jones 
interaction and 
the dipole-dipole interaction, respectively.  $\epsilon$ and $\sigma$ 
are the Lennard-Jones parameters.  
Since the total dipole mement of each molecule depends on the dipole moments 
of other molecules, the energy is calculated by an iterative procedure.  
Details of the iterative procedure are discussed in the next section.

\section{Simulation Details}

	Throughout this study, data are presented in the reduced units, 
denoted by the superscript (*).  
The units of energy and length are 
reduced by the Lennard-Jones parameters $\epsilon$ and $\sigma$, respectively.  
Reduced temperature is $T^*=kT/\epsilon$.  
Dipole moment ($\bbox{m}$) and polarizability 
($\bbox{\alpha}$)
are reduced by the Lennard-Jones parameters as 
$\displaystyle\bbox{m}^*=\bbox{m}/\sqrt{\epsilon \sigma^3}$ and 
$\bbox{\alpha}^*=\bbox{\alpha}/\sigma^3$.
We study Stockmayer fluids with permanent dipole moments 
$|\bbox{m_0}^*|=$ 1.0 and 2.0 
and isotropic polarizabilities $\alpha^*=$ 0.00, 0.03, and 0.06.  

	In GCMC simulations, 
at each Monte Carlo step, a new microstate is generated by a displacement, 
rotation, and creation or destruction of a molecule. 
The state thus generated is probabilistically accepted 
so that the limiting distribution of 
generated microstates obeys the grand canonical ensemble.  We use 
10\% of the Monte Carlo moves for displacement, 10\% for rotation, 
40\% for creation, and 40\% for destruction.  In the energy calculation 
of polarizable models, the total interaction of the system is
calculated by an iterative procedure \cite{patey,vesely,levy} 
described below.  

The initial values of properties for a configuration are indicated by (0), and 
the estimates for those properties at the k-th iteration by (k).  
	When a molecule is either displaced, rotated, created or destroyed, 
the initial estimate of the electric field at the center of molecule $i$ is 

\begin{equation}
	\bbox{E}_i(1) = \sum_{j\neq i} \bbox{T}_{ij} \cdot \bbox{m}_j(0) 
\end{equation}

\noindent Then, the first estimate of the total dipole mement of molecule i is 

\begin{equation}
	\bbox{m}_i(1)= \bbox{m}_{0,i}+\bbox{\alpha} \cdot 
\bbox{E}_i(1), 
\end{equation}

\noindent and the total electrostatic interaction is 

\begin{equation}
	U_{dp}(1) = -\frac{1}{2}\sum_i \bbox{m}_{0,i} \cdot \bbox{E}_i(1) .
\end{equation}

This way, the k-th estimates of the electric field, the dipole mement, and 
the dipole-dipole interaction are 

\begin{eqnarray}
	\bbox{E}_i(k) = \sum_{j\neq i} \bbox{T}_{ij} \cdot \bbox{m}_j(k-1) 
\nonumber\\
	\bbox{m}_i(k)= \bbox{m}_{0,i}+\bbox{\alpha} \cdot \bbox{E}_i(k), 
\nonumber\\
	U_{dp}(k) = -\frac{1}{2}\sum_i \bbox{m}_{0,i} \cdot \bbox{E}_i(k)
\end{eqnarray}

\noindent respectively.  This procedure is repeated until $U_{dp}(k)$ 
is converged 
so that $|U_{dp}(k)-U_{dp}(k-1)|/|U_{dp}(k-1)| < 0.0001$ for two consecutive 
iterations.  Typically, the 
number of iterations required for this criterion at each Monte Carlo move 
turns out to be 
3 to 4 for polarizability of $\alpha^*=$ 0.03 and 3 to 6 for 
$\alpha^*=$ 0.06 (see tables 1 and 2).  The initial value of 
dipole moment of molecule $i(\bbox{m}_i(0))$ at each Monte Carlo
step is chosen to be the dipole moment before the move, except for
a newly inserted molecule for which the dipole moment is chosen to
be the same as the permanent dipole moment.  The dipole points to
a random direction for a newly inserted molecule.
In order to minimize the size effect for the dipole-dipole interaction, 
we use the Ewald sum with 256 vectors for the reciprocal space terms 
\cite{deleeuw} 
for the model with $|\bbox{m_0}^*|=$ 1.0 and $\alpha^*=$ 0.00, 
and 514 vectors for the other models that we study.  
Approximate overall cpu-time per Monte Carlo step for 
the polarizable models is 2 times slower for $\alpha^*=$ 0.03 and 
2.4 times slower for $\alpha^*=$ 0.06
than the cpu-time for $\alpha^*=$ 0.00 by our code 
designed for polarizable 
models with the Ewald sum.  
Simulations of non-polarizable models by our code 
for polarizable models are, in turn, approximately 4 times slower than those by our code 
specifically designed for non-polarizable models, because the energy 
calculation of polarizable models requires the calculation of 
the local electric field of each molecule.  

When two polarizable molecules approach unphysically close to each other
during simulation, the electrostatic attraction 
increases faster than the repulsive part of the Lennard-Jones potential 
due to the increasing magnitude of the total dipoles and eventually 
the two molecules 
overlap \cite{levy}.  In order to avoid this effect, we set  
a saturation point of the total dipole equal to twice the magnitude of 
the permanent dipole.  
This treatment seems to be reasonable because the average of 
the total dipole moment 
is far smaller than twice the magnitude of the permanent dipole in each of the 
simulations we performed (see tables 1 and 2).  
For the Lennard-Jones interaction, the potential is cut off at half of the 
simulation box length and the standard long range correction is 
applied \cite{at}.  

	During the simulations, the number of observations 
of a given number of particles and energy is counted and the two dimensional 
histogram $f_{\mu,V,T}(N,U_N)$ is made.  The grid of the histogram 
for energy is 
chosen to be 0.01 in the reduced unit. The two dimensional histograms 
of the number of particles and the energy are analyzed to obtain 
coexistence properties for a range of temperature by the method described 
in section 2.  

In most simulations, the volume is chosen to be $V^*$=216.  For simulations of 
low densities, the volume is chosen to be $V^*$=2160 so that the simulation 
box can accommodate a large enough number of particles to measure 
a density difference 
as small as $\Delta \rho ^* = \Delta [1/V^*] \simeq $ 0.0005. 
In combining the histograms of simulations of different volumes, we rescale 
the histogram for $V^*$=2160 to that for $V^*$=216 assuming 

\begin{equation}
	\Omega(N, V, T) = {\Omega(k N,k V,T)}^{\frac{1}{k}}, 
\end{equation}
or, in terms of the component of the histogram,
\begin{equation}
	f_{\mu,V,T}(N,U_N) = {f_{\mu,kV,T}(k N, U_{k N})}^{\frac{1}{k}}
\end{equation}

\noindent which is a reasonable assumption away from the critical point, since
the logarithms of the microcanonical and canonical partition functions are
extensive properties of the system.  

To fix the simulation specific constant $C$, we use the pressure calculated by 
the virial theorem in one of the simulations in the gas phase.  
The derivative of 
energy in terms of volume is calculated by an approximation, 

\begin{equation}
	\frac{dU}{dV} \simeq \frac{U(V+\delta V)-U(V)}{\delta V} .
\end{equation}

\noindent The $\delta V$ is chosen to be about 3\% of the volume $V$.  The energy $U(V+\delta V)$ 
is calculated every time a Monte Carlo move is accepted.  

For each model, we first estimate the approximate location 
of the critical point 
by looking up literature values for similar models \cite{smit1,smit2} 
and by performing several short test simulations.  
Then, we perform a series of GCMC simulations at a temperature slightly 
above the 
critical temperature for various chemical potentials to sample a wide 
range of density.  
The reasons why we choose 
a temperature slightly above the critical temperature are that large 
density fluctuations near the critical point make it easier to sample a wide 
range of density in a single simulation and that for subcritical temperatures
the system tends to fluctuate infrequently between the gas and liquid 
densities, 
making sampling of both phases difficult.   

An example is illustrated in figure 1 for the model with 
$|\bbox{m_0}^*|=1.0 $ and $ \alpha ^*=0.03$.  
At $T^*$=1.5, we performed GCMC simulations with 
various chemical potentials ($\mu ^*= -7.00 \sim -1.00$) to 
cover the density range of interest.  
The mean densities 
calculated from these simulations are shown in figure 1 by circles.
The coexistence properties for a range of temperatures ($T^*=1.0 \sim 1.5$ 
for this example)
are calculated from these simulation results.  

In order to make sure that the simulations are sampling the configurations 
relevant to the phase equilibrium properties, we performed a new series of 
simulations with temperatures and chemical potentials that are near the 
corresponding properties at phase coexistence for temperatures lower 
than the temperature for the first simulations ($T^*$=1.0, 1.1, 1.2, and 1.3).  
The chemical potentials for these simulations are chosen to be 
the values at coexistence estimated from the first simulation results 
for each temperature.
The mean densities calculated from the 
new series of GCMC simulations 
are shown in figure 1 by squares.  

The histograms obtained from the new series 
of simulations (below the critical point) 
and four of the first simulations (near the critical point) 
were analyzed to calculate the phase coexistence properties.  
The distributions of density from the histograms used for analysis 
are shown in figure 2 and the conditions of the simulations are listed 
in table 1.
The fitting of the calculated coexistence densities to 
the law of rectilinear diameters and a scaling law, 
assuming that dipolar fluids obey the Ising exponent ($\beta=0.326$), 
is shown in figure 1 by dashed line.  
We performed five sets of simulations and analyses. 
The length of each GCMC simulation was 1 million Monte Carlo steps. 
The averages and the 
error-bars were calculated from the five sets of data.  For error-bars, 
we use the square root of the variance of mean.  Thus calculated error-bars 
are not, in a strict sense, statistical errors 
because there are several sources 
of statistical errors that propagate to the final results: for example, 
the thermodynamic states of interest here are not sampled 
an exactly equal number 
of times in the simulations.  However, 
they are still good estimates of the reliability of the simulations and the 
data analysis.  

	As we can see in figure 1, 
the mean densities of the second series of simulations (squares) 
agree well with the final results (dashed line).
Since the chemical potentials in the second series of simulations 
are based on the estimates from the first simulations (circles), 
it turns out that the estimates of coexistence densities from the 
first simulations are fairly accurate, considering that all the 
simulations were performed at one temperature ($T^*=1.5$).  

\section{Results}

	The magnitude of the average total dipole moment and 
the number of iterations necessary for the convergence of 
the energy calculation, with the criterion given in the previous 
section, are listed for polarizable models along with the 
conditions used for the original GCMC simulations in tables 1 and 2.
The number of iterations necessary for energy calculation 
turns out to be 3 to 6, depending on the models and the thermodynamic 
states.  
The calculated magnitude of the induced dipole is larger for the models with 
larger permanent dipoles and larger polarizabilities.  For the model with 
$|\bbox{m_0}^*|=2.0$ and $\alpha ^*=0.06$, the induced dipole is 
as large as 30 \% 
of the permanent dipole at $T^*=1.0$ and $\rho^* \simeq 0.8$.    

Examples of the probability density distributions from the GCMC simulations are shown in 
figure 2  for the Stockmayer fluid with $|\bbox{m_0}^*|=1.0$ and 
$\alpha ^*=0.03$.
By reweighting and combining the histograms obtained from the 
simulations, we get the density distribution for any temperature 
and chemical potential.   
Examples of the density distributions at coexistence are shown in figure 3 for 
the Stockmayer fluid with $|\bbox{m_0}^*|=1.0$ and $\alpha ^*=0.03$.   
Because of the small system size of the simulations, 
the two peaks in the density distribution at coexistence overlap far 
below the critical temperature.  
We calculate the coexistence properties up to 
the temperature where the two peaks start to overlap.  
The results are shown in figures 4 to 9,  and tables 3, 4 and 5.
The critical temperature increases for the models of higher dipole
moment and higher polarizability.
The heat of vaporization is calculated from the results of 
internal energy, pressure, and density at coexistence.

We estimate the approximate values of the infinite-system critical 
temperature and density 
by fitting the simulation results to the law of rectilinear 
diameters and a scaling law, assuming that dipolar fluids obey 
the Ising exponent ($\beta=0.326$).  Finite-size scaling
methods\cite{wilding} can be used to locate the critical point with a much
higher accuracy than the present study, but require a series of
simulations for different system sizes.  
The critical temperature increases 
as the polarizability increases 
for both dipoles ($|\bbox{m_0}^*|=1.0$ and $2.0$) 
that we studied (see table 5).  The effect of polarizability on the
critical density is not pronounced.  Our results seem to indicate
a slight increase in the critical density at higher polarizabilities,
but the differences are comparable to the statistical uncertainties
of the calculations.

	The calculated coexistence density, pressure, and heat of 
vaporization are compared with the  
renormalization perturbation theory by Wertheim figures 4 to 9.  
For the theoretical calculation, we follow the prescription given by 
Venkatasubramanian et al. \cite{gubbins1} and use 
the equation of state by Johnson et al. \cite{johnson} and the coefficients 
of the pair and triplet correlation functions calculated by 
Flytzani-Stephanopoulos et al. \cite{gubbins2} and Gubbins and Twu 
\cite{gubbins3} for the Lennard-Jones reference fluid.  
The agreement between simulation and theory 
is relatively good for $|\bbox{m_0^*}|=1.0$, but poor for 
$|\bbox{m_0^*}|=2.0$.  Statistical uncertainties of the results
are quite small, confirming the computational advantages of the histogram 
reweighting method.

\section{Conclusions}
We have calculated the phase coexistence properties of polarizable and 
non-polarizable Stockmayer fluids by 
the GCMC histogram reweighting method.  In the histogram reweighting method,
the grand canonical partition function 
of the system is constructed, from which any thermodynamic property 
can be derived by statistical thermodynamic analysis.  
Since the thermodynamic state for 
the grand canonical partition function can be continuously changed 
by ``reweighting'' the histograms, thermodynamic properties at 
thermodynamic states that are different from the thermodynamic state 
of the original simulation can be calculated.  The statistical
uncertainties of the calculated results are small, confirming the computational advantages of the histogram 
reweighting method over existing methods (such as Gibbs ensemble Monte
Carlo\cite{panagiot}).

Our results for the coexistence properties were compared to 
Wertheim's renormalization perturbation
theory.  Differences between theoretical and simulation results
are within 2 \% for the smaller dipole ($|\bbox{m_0^*}|=1.0$) 
but only within 10 \% 
for the higher dipole ($|\bbox{m_0^*}|=2.0$) that we studied.

\section{Acknowledgments}
The authors would like to thank Mr. Gerassimos Orkoulas 
for bringing the histogram reweighting method to their attention 
and Dr. Thomas Kraska for helpful discussions on Wertheim's theory.  
Financial support for this work was provided 
by the Department of Energy, Office of Basic Energy Science. 
Supercomputing time was provided by the Cornell Theory Center.

\begin{table}
\caption{Conditions for the GCMC simulations 
(temperature $T^*$, chemical potenitial $\mu ^*$, and volume $V^*$) 
for polarizable Stockmayer fluids
with $|\bbox{m_0^*}|=1.0$ and different polarizabilities ($\alpha ^*$).
The calculated average density ($\rho ^*$) 
, average magnitude of the total dipole ($|\bbox{m}^*|$) and 
the average number of iterations ($k_{itr}$) necessary for calculation of 
the energy for polarizable models 
at each Monte Carlo step (see text) 
are also listed.  
The numbers in parentheses are statistical uncertanties, in units of
the last decimal point listed.}
\begin{tabular}{c|c|l|r|r|l|c}
 $\alpha ^*$  &   $T^*$  &  \ \ $\mu^*$ & $V^*$  & $ \rho^*$ \ \ \ \ \ & \ \ \ $|\bbox{m}^*|$ 
&	$k_{itr}$ \\
\hline
0.03   &  1.00       &  -4.60      &  2160     &  0.0118(000)     &  1.0023(00)  & 3.2 	\\ 
 	   &  1.10       &  -4.43      &  2160     &  0.0231(001)     &  1.0039(01)  & 3.3   	\\
       &  1.20       &  -4.30       &  216     &  0.0410(001)     &  1.0051(00)  & 3.8   	\\
       &  1.30       &  -4.17       &  216     &  0.0732(007)     &  1.0084(01)  & 3.8   	\\
       &  1.50       &  -4.20       &  216     &  0.1246(018)     &  1.0118(02)  & 4.0   	\\
       &  1.50       &  -4.00       &  216     &  0.2248(087)     &  1.0199(07)  & 4.0   	\\
       &  1.50       &  -3.90       &  216     &  0.4118(118)     &  1.0335(08)  & 4.1   	\\
           &  1.50       &  -3.72       &  216     &  0.5225(068)     &  1.0410(07)  & 4.1   	\\
           &  1.30       &  -4.07       &  216     &  0.6000(060)     &  1.0507(06)  & 4.0   	\\
           &  1.20       &  -4.20       &  216     &  0.6679(030)     &  1.0572(04)  & 4.0   	\\
           &  1.10       &  -4.33       &  216     &  0.7306(055)     &  1.0646(08)  & 4.0   	\\
           &  1.00       &  -4.50       &  216     &  0.7674(024)     &  1.0699(05)  & 4.0  	\\
\hline
  0.06     &  1.00       &  -4.70      &  2160     &  0.0106(000)     &  1.0048(01)   & 3.4  	\\
          &  1.10       &  -4.53      &  2160     &  0.0209(001)     &  1.0086(01)   & 3.6  	\\
          &  1.20       &  -4.40       &  216     &  0.0370(002)     &  1.0109(02)   & 4.5  	\\
          &  1.30       &  -4.17       &  216     &  0.0806(004)     &  1.0224(02)   & 4.6  	\\
          &  1.50       &  -4.10       &  216     &  0.2920(268)     &  1.0607(48)  & 5.2  	\\
          &  1.50       &  -4.085      &  216     &  0.3074(282)     &  1.0638(55)  & 5.2  	\\
          &  1.40       &  -4.03       &  216     &  0.6005(052)     &  1.1213(17)  & 5.7  	\\
          &  1.30       &  -4.07       &  216     &  0.6666(047)     &  1.1364(15)  & 5.7  	\\
          &  1.20       &  -4.20       &  216     &  0.7217(043)     &  1.1530(16)  & 5.7   	\\
          &  1.10       &  -4.33       &  216     &  0.7819(046)     &  1.1714(14)  & 5.7  	\\
          &  1.00       &  -4.50       &  216     &  0.8057(061)     &  1.1844(24)  & 5.7  	\\
\end{tabular}
\end{table}
\begin{table}
\caption{Conditions for the GCMC simulations 
with the calculated average density, average magnitude of the total dipole
and the number of iteration necessary for calculation of energy 
for polarizable Stockmayer fluids
with $|\bbox{m_0}^*|=2.0$ and different polarizabilities.  
Notation is the same as for Table 1.}
\begin{tabular}{c|c|l|r|r|l|c}
   $\alpha ^*$ &  $T^*$ & \ \ $\mu^*$ & $V^*$  & $ \rho^*$ \ \ \ \ \ & \ \ \ $|\bbox{m}^*|$ 
&	$k_{itr}$ \\
\hline
  0.03     &  1.70       &  -7.90      &  2160     &  0.0129(001)     &  2.0177(004) & 3.2  	\\
          &  1.80       &  -7.70      &  2160     &  0.0211(002)     &  2.0264(008) & 3.2  	\\
          &  1.90       &  -7.52      &  2160     &  0.0327(004)     &  2.0340(010) & 3.2  	\\
          &  2.00       &  -7.37       &  216     &  0.0497(010)     &  2.0368(011) & 3.7  	\\
          &  2.10       &  -7.24       &  216     &  0.0813(036)     &  2.0513(019) & 3.6   	\\
          &  2.20       &  -7.16       &  216     &  0.1305(180)     &  2.0669(047) & 3.7   	\\
          &  2.40       &  -7.00       &  216     &  0.2128(130)     &  2.0897(037) & 3.7   	\\
          &  2.40       &  -6.75       &  216     &  0.3724(158)     &  2.1275(041) & 3.7   	\\
          &  2.20       &  -6.96       &  216     &  0.5143(096)     &  2.1641(022) & 3.8   	\\
          &  2.10       &  -7.14       &  216     &  0.5723(107)     &  2.1793(025) & 3.8   	\\
          &  2.00       &  -7.27       &  216     &  0.6289(090)     &  2.1918(020) & 3.8   	\\
          &  1.90       &  -7.42       &  216     &  0.6787(037)     &  2.2046(010) & 3.8   	\\
          &  1.80       &  -7.60       &  216     &  0.7147(071)     &  2.2158(019) & 3.8   	\\
          &  1.70       &  -7.80       &  216     &  0.7540(059)     &  2.2255(012) & 3.8  	\\
\hline
  0.06     &  2.00       &  -8.71      &  2160     &  0.0176(002)     &  2.0443(016) & 3.4   	\\
          &  2.10       &  -8.51      &  2160     &  0.0260(001)     &  2.0579(007)  & 3.4   	\\
          &  2.20       &  -8.32      &  2160     &  0.0376(003)     &  2.0766(018) & 3.5   	\\
          &  2.30       &  -8.14       &  216     &  0.0555(012)     &  2.0869(030) & 4.4   	\\
          &  2.40       &  -7.98       &  216     &  0.0900(109)     &  2.1219(114) & 4.5   	\\
          &  2.50       &  -7.85       &  216     &  0.1378(136)     &  2.1719(149) & 4.6   	\\
          &  2.70       &  -7.60       &  216     &  0.2940(184)     &  2.2762(121) & 4.8   	\\
          &  2.70       &  -7.40       &  216     &  0.4318(182)     &  2.3574(124) & 5.0   	\\
          &  2.50       &  -7.70       &  216     &  0.5335(080)     &  2.4353(063)  & 5.1   	\\
          &  2.40       &  -7.83       &  216     &  0.5889(059)     &  2.4725(046)  & 5.1   	\\
          &  2.30       &  -8.04       &  216     &  0.6564(109)     &  2.5199(068)  & 5.2   	\\
          &  2.20       &  -8.22       &  216     &  0.6927(068)     &  2.5417(044)  & 5.2   	\\
          &  2.10       &  -8.41       &  216     &  0.7284(109)     &  2.5724(076)  & 5.2   	\\
          &  2.00       &  -8.61       &  216     &  0.7652(084)     &  2.6034(066)  & 5.2  	\\
\end{tabular}
\end{table}
\begin{table}
\squeezetable
\caption{Calculated coexistence properties: chemical potential ($\mu ^*$), 
pressure ($p^*$), gas phase density ($\rho ^*_g$), liquid 
phase density ($\rho ^*_l$), internal energies per molecule of 
gas phase ($u^*_g$) 
and of liquid phase ($u^*_l$), heat of vaporization per molecule ($\Delta h^*$) 
at different temperatures ($T^*$) for polarizable Stockmayer fluids 
with $|\bbox{m_0}^*|=1.0$ and $\alpha ^*=$ 0.00, 0.03, and 0.06.  
The numbers in parentheses are statistical uncertanties, in units of
the last decimal point listed.}
\begin{tabular}{c|c|r|r|r|r|r|r|r}
 $\alpha ^*$  &  $T^*$ & $\mu^*$ \ \ \ \ & $p^*$ \ \ \ \ 
 & $\rho^*_g$ \ \ \ \ & $\rho^*_l$ \ \ \ \ 
 & $u^*_g$ \ \ \ \ & $u^*_l$ \ \ \ \
 & $\Delta h^*$ \ \ \ \\
 \hline
 0.00&    1.00   &  -4.409(4) &   .0179(1) &   .0195(02) &    .754(3) &    -.276(05) &   -6.11(3) &    6.72(3)\\
    &	1.05   &  -4.320(3) &   .0253(1) &   .0258(02) &    .732(2) &    -.338(04) &   -5.88(2) &    6.49(2)\\
    &	1.10   &  -4.240(3) &   .0345(1) &   .0343(02) &    .708(1) &    -.407(03) &   -5.65(1) &    6.20(2)\\
    &	1.15   &  -4.167(2) &   .0456(1) &   .0451(02) &    .682(2) &    -.480(03) &   -5.41(2) &    5.87(2)\\
    &	1.20   &  -4.102(2) &   .0590(1) &   .0587(01) &    .648(2) &    -.568(02) &   -5.11(2) &    5.45(2)\\
    &	1.25   &  -4.045(2) &   .0748(1) &   .0766(01) &    .612(2) &    -.696(01) &   -4.79(1) &    4.95(1)\\
    &	1.30   &  -3.994(2) &   .0934(1) &   .1017(03) &    .570(1) &    -.891(02) &   -4.44(1) &    4.30(1)\\
 \hline
    0.03 & 1.00   &  -4.565(2) &   .0138(1) &   .0162(01) &    .772(1) &    -.259(02) &   -6.41(1) &    6.99(1)\\
    &	1.05   &  -4.470(3) &   .0203(1) &   .0219(01) &    .751(3) &    -.304(03) &   -6.19(2) &    6.79(2)\\
    &	1.10   &  -4.384(3) &   .0284(1) &   .0291(01) &    .729(2) &    -.349(03) &   -5.96(2) &    6.55(2)\\
    &	1.15   &  -4.306(4) &   .0384(1) &   .0381(02) &    .700(1) &    -.400(03) &   -5.68(1) &    6.24(1)\\
    &	1.20   &  -4.237(3) &   .0503(2) &   .0497(02) &    .668(3) &    -.477(03) &   -5.39(2) &    5.85(2)\\
    &	1.25   &  -4.176(3) &   .0645(2) &   .0650(03) &    .636(3) &    -.596(03) &   -5.10(2) &    5.40(2)\\
    &	1.30   &  -4.121(3) &   .0813(2) &   .0857(04) &    .601(3) &    -.770(03) &   -4.80(2) &    4.84(3)\\
    &	1.35   &  -4.072(2) &   .1011(2) &   .1150(05) &    .556(2) &   -1.016(04) &   -4.42(2) &    4.10(2)\\
 \hline
    0.06 & 1.00   &  -4.773(3) &   .0105(1) &   .0135(00) &    .797(4) &    -.241(04) &   -6.83(3) &    7.35(3)\\
    &	1.05   &  -4.669(3) &   .0162(1) &   .0182(01) &    .777(2) &    -.281(05) &   -6.61(2) &    7.19(2)\\
    &	1.10   &  -4.574(2) &   .0233(1) &   .0241(01) &    .755(3) &    -.320(05) &   -6.37(2) &    6.98(3)\\
    &	1.15   &  -4.487(2) &   .0320(1) &   .0315(01) &    .729(2) &    -.358(05) &   -6.11(1) &    6.73(2)\\
    &	1.20   &  -4.410(2) &   .0426(2) &   .0408(01) &    .701(1) &    -.411(03) &   -5.83(1) &    6.41(1)\\
    &	1.25   &  -4.341(2) &   .0551(2) &   .0528(01) &    .668(2) &    -.498(01) &   -5.52(2) &    5.99(2)\\
    &	1.30   &  -4.279(3) &   .0699(3) &   .0688(04) &    .636(2) &    -.634(03) &   -5.23(2) &    5.50(2)\\
    &	1.35   &  -4.224(4) &   .0874(4) &   .0908(09) &    .602(2) &    -.828(09) &   -4.92(1) &    4.91(2)\\
    &	1.40   &  -4.174(4) &   .1081(6) &   .1218(23) &    .555(4) &   -1.097(22) &   -4.52(3) &    4.11(4)\\
\end{tabular}
\end{table}
\begin{table}
\squeezetable
\caption{Calculated coexistence properties for polarizable Stockmayer fluids 
with $|\bbox{m_0}^*|=2.0$ and $\alpha ^*=$ 0.00, 0.03, and 0.06.  
Notation is the same as in table 3.}

\begin{tabular}{c|c|r r r r r r r}
 $\alpha ^* $ & $T^*$ & $\mu^*$ \ \ \ \ & $p^*$ \ \ \ \ 
 & $\rho^*_g$ \ \ \ \ & $\rho^*_l$ \ \ \ \ 
 & $u^*_g$ \ \ \ \ & $u^*_l$ \ \ \ \
 & $\Delta h^*$ \ \ \ \\
 \hline
 0.00 &   1.60   &  -7.177(3) &   .0224(0) &   .0222(01) &    .726(2) &   -1.16(1) &  -10.16(2) &    9.97(2)\\
 &   1.65   &  -7.078(4) &   .0300(1) &   .0273(02) &    .706(2) &   -1.28(1) &   -9.89(3) &    9.67(3)\\
 &   1.70   &  -6.986(4) &   .0388(1) &   .0335(02) &    .682(2) &   -1.40(1) &   -9.58(3) &    9.28(4)\\
 &   1.75   &  -6.900(4) &   .0490(1) &   .0411(03) &    .654(2) &   -1.53(1) &   -9.22(3) &    8.82(3)\\
 &   1.80   &  -6.822(4) &   .0607(2) &   .0507(04) &    .626(2) &   -1.67(1) &   -8.87(2) &    8.30(3)\\
 &   1.85   &  -6.750(4) &   .0741(3) &   .0634(05) &    .599(2) &   -1.86(1) &   -8.53(2) &    7.72(3)\\
 &   1.90   &  -6.683(4) &   .0896(3) &   .0811(07) &    .569(4) &   -2.13(2) &   -8.17(4) &    6.98(4)\\
 \hline
 0.03 &   1.70   &  -7.893(6) &   .0233(4) &   .0183(01) &    .758(4) &   -1.11(1) &  -11.48(6) &   11.62(9)\\
 &   1.75   &  -7.783(4) &   .0301(4) &   .0222(01) &    .741(5) &   -1.22(1) &  -11.21(6) &   11.30(8)\\
 &   1.80   &  -7.679(3) &   .0379(4) &   .0270(01) &    .720(6) &   -1.34(1) &  -10.91(7) &   10.91(8)\\
 &   1.85   &  -7.582(2) &   .0469(4) &   .0327(00) &    .697(5) &   -1.46(1) &  -10.58(6) &   10.48(8)\\
 &   1.90   &  -7.492(3) &   .0572(5) &   .0396(00) &    .674(4) &   -1.58(1) &  -10.25(5) &   10.02(6)\\
 &   1.95   &  -7.407(3) &   .0690(5) &   .0480(01) &    .650(4) &   -1.71(1) &   -9.92(4) &    9.53(5)\\
 &   2.00   &  -7.329(4) &   .0823(5) &   .0587(02) &    .624(4) &   -1.87(0) &   -9.56(4) &    8.95(5)\\
 &   2.05   &  -7.255(4) &   .0974(6) &   .0730(05) &    .593(4) &   -2.10(1) &   -9.16(4) &    8.21(5)\\
 &   2.10   &  -7.187(5) &   .1146(7) &   .0927(11) &    .556(4) &   -2.44(2) &   -8.69(4) &    7.27(6)\\
 \hline
 0.06 &   2.00   &  -8.695(8) &   .0357(4) &   .0232(02) &    .761(1) &   -1.19(2) &  -13.00(3) &   13.30(4)\\
 &   2.05   &  -8.581(8) &   .0440(4) &   .0275(02) &    .749(1) &   -1.31(2) &  -12.75(4) &   12.98(5)\\
 &   2.10   &  -8.471(8) &   .0535(4) &   .0325(02) &    .732(3) &   -1.44(2) &  -12.44(5) &   12.57(5)\\
 &   2.15   &  -8.367(8) &   .0641(4) &   .0385(02) &    .709(4) &   -1.58(2) &  -12.07(4) &   12.07(5)\\
 &   2.20   &  -8.270(7) &   .0761(4) &   .0455(03) &    .686(4) &   -1.71(2) &  -11.69(4) &   11.54(5)\\
 &   2.25   &  -8.178(7) &   .0895(4) &   .0539(04) &    .663(3) &   -1.86(2) &  -11.33(4) &   11.00(5)\\
 &   2.30   &  -8.092(6) &   .1044(4) &   .0644(05) &    .639(4) &   -2.05(2) &  -10.95(5) &   10.36(6)\\
 &   2.35   &  -8.010(6) &   .1211(5) &   .0784(07) &    .609(5) &   -2.30(2) &  -10.51(7) &    9.55(8)\\
 &   2.40   &  -7.934(5) &   .1397(6) &   .0969(12) &    .573(7) &   -2.66(2) &   -9.99(9) &    8.53(9)\\
\end{tabular}
\end{table}
\clearpage
\begin{table}
\caption{Estimates of critical temperature and density for 
polarizable Stockmayer fluids 
as a function of the permanent dipole ($|\bbox{m_0}^*|$) 
and polarizability ($\alpha ^*$).}
\begin{tabular}{c|c|l l}
 $|\bbox{m_0}^*|$  &  $\alpha ^*$  & \ \ $T_{cr}^*$  & \ \ $\rho _{cr}^*$ \\
\hline
1.00	&	0.00	&	1.400(3)	&	0.318(3) \\
    	&	0.03	&	1.432(6)	&	0.322(5) \\
    	&	0.06	&	1.478(8)	&	0.328(6) \\
\hline
2.00	&	0.00	&	2.05(1)	&	0.306(08) \\
    	&	0.03	&	2.22(1)	&	0.302(09) \\
    	&	0.06	&	2.53(2)	&	0.312(16) \\
\end{tabular}
\end{table}

\input epsf
\begin{figure}
\epsfxsize=3.3in
\epsffile{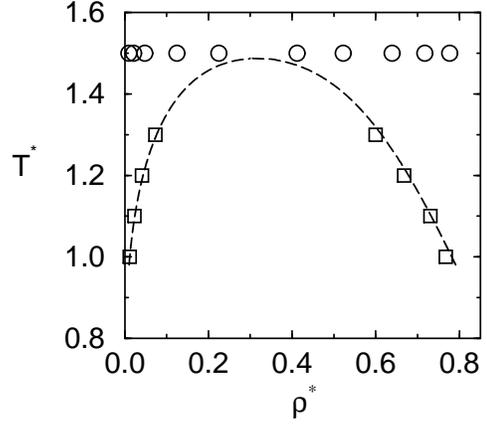}
\caption{Temperature and mean density of example GCMC simulations.  The initial 
simulations are performed to cover a wide range of density 
at a temperature slightly higher than 
the estimated critical point ($T^*$=1.5) with chemical potentials 
of $\mu ^*$=-7.00, -6.00, -5.00, -4.20, -4.00, -3.90, -3.72, 
-3.00, -2.00, and -1.00 (circles, from left to right).  The second 
series of simulations are performed at temperatures and chemical potentials 
for phase coexistence (squares, see also table 1),  as calculated from the initial 
simulations.  Dashed line is the fitting of the coexistence curve to the results 
from the \emph{second} series of simulations.}
\end{figure}

\begin{figure}
\epsfxsize=3.3in
\epsffile{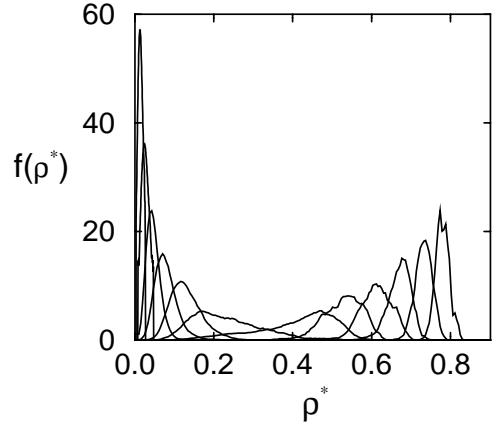}
\caption{Density distributions from the GCMC simulations 
with $|\bbox{m_0}^*|=1.0$ and $\alpha ^*=0.03$.  The chemical potential 
and the temperature for each simulation is, from left to right, 
($\mu ^*$= -4.6, $T^*$=1.0), (-4.43, 1.1), (-4.30,1.2), (-4.17,1.3), 
(-4.20,1.5), (-4.00,1.5), (-3.90,1.5), (-3.72,1.5), (-4.07,1.3), 
(-4.20,1.2), (-4.33,1.1), (-4.50,1.0) (see also table 1).  The whole 
density range of interest is sampled by these simulations.  
}
\end{figure}

\begin{figure}
\epsfxsize=3.3in
\epsffile{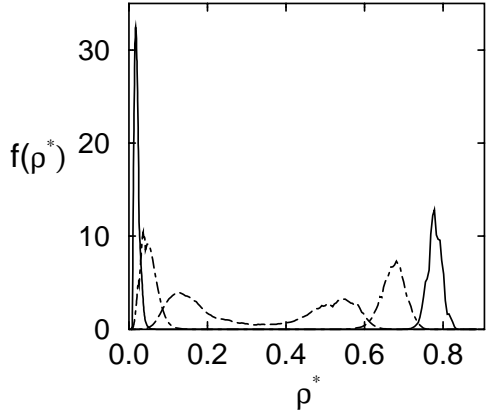}
\caption{Density distributions at two-phase coexistence for the 
polarizable Stockmayer fluid 
with $|\bbox{m_0}^*|=1.0$ and $\alpha ^*=0.03$, calculated by 
reweighting the histogram obtained from the GCMC simulations (see table 1 
and figure 1).  The distributions are shown for 
$T^*=$ 1.00 (solid line), 1.20 (dot-dashed line), and 1.40 (dashed line).  
 }
\end{figure}

\begin{figure}
\epsfxsize=3.3in
\epsffile{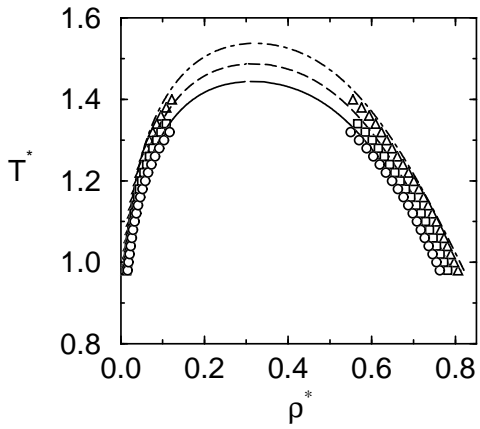}
\caption{Coexistence densities for the polarizable Stockmayer fluids 
with $|\bbox{m_0}^*|=1.0$.
Circles, squares, triangles are for $\alpha ^*=$ 0.00, 0.03, and 0.06, 
respectively.  
The error bars are about the same or smaller than the size of the symbols.  
Solid line, dashed line, and dot-dashed line are 
the results of the Wertheim's perturbation theory 
for $\alpha ^*=$ 0.00, 0.03, and 0.06 respectively.
 }
\end{figure}

\begin{figure}
\epsfxsize=3.3in
\epsffile{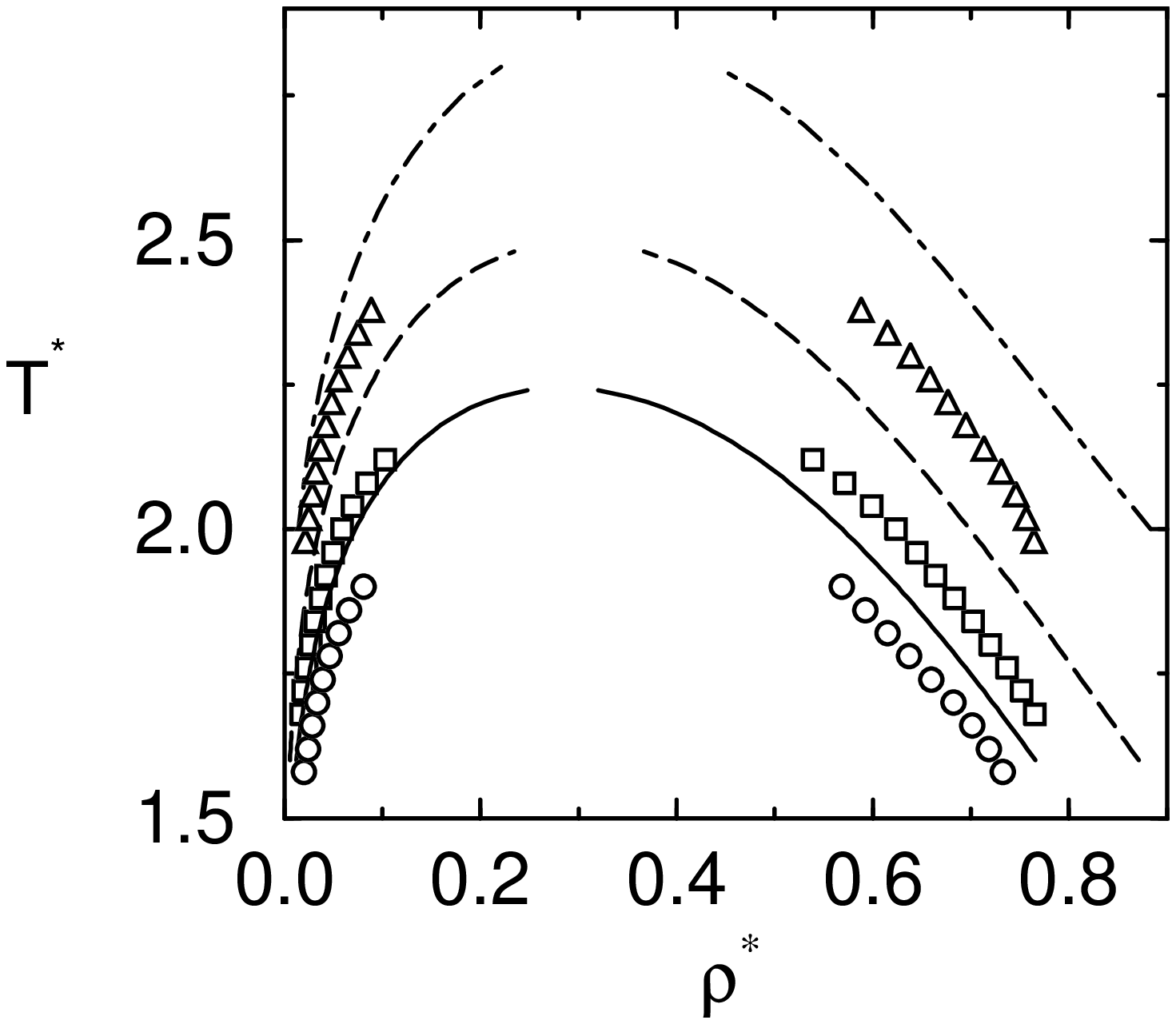}
\caption{Coexistence densities for the polarizable Stockmayer fluids 
with $|\bbox{m_0}^*|=2.0$.
The notation is the same as in figure 4.  
 }
\end{figure}

\begin{figure}
\epsfxsize=3.3in
\epsffile{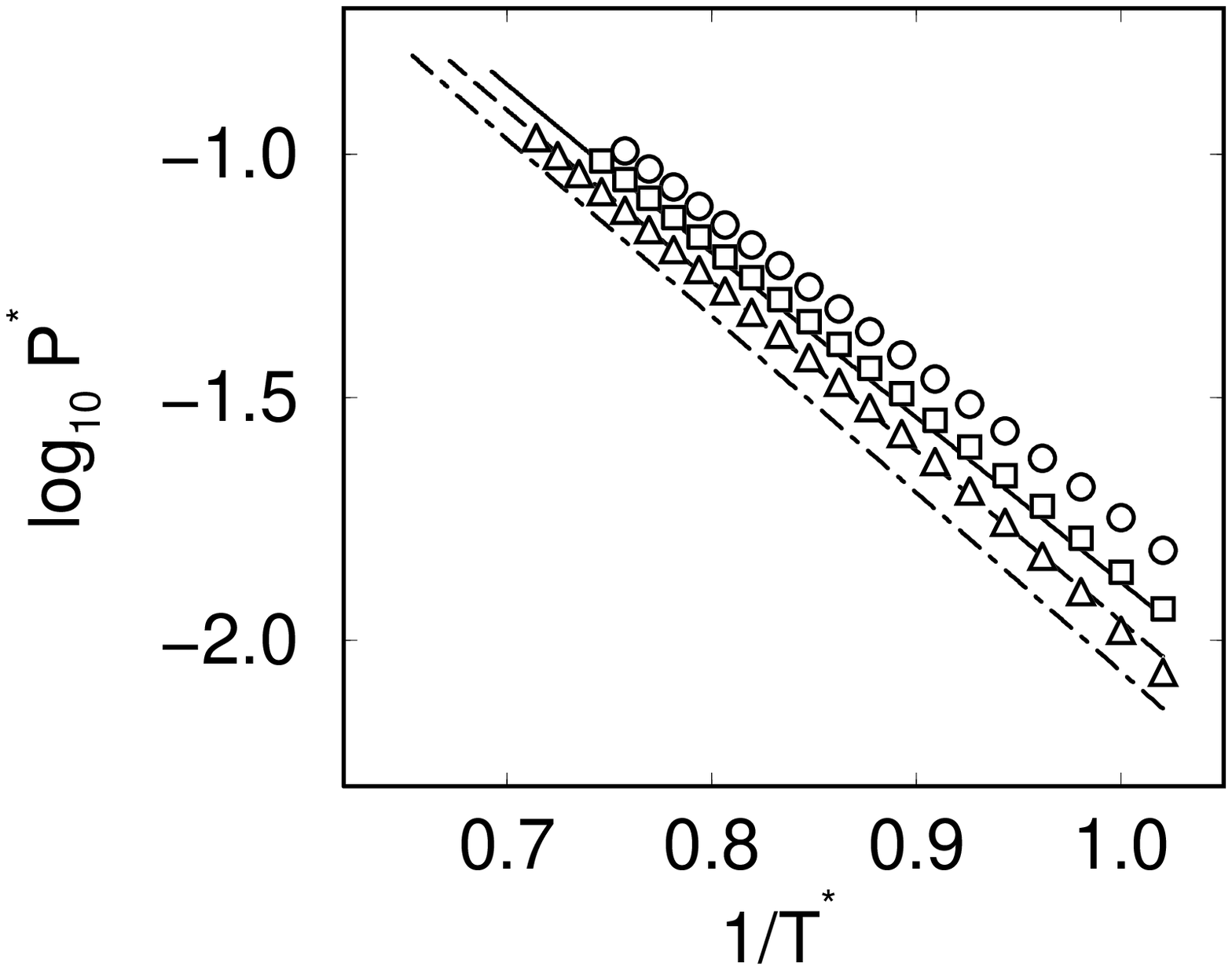}
\caption{Coexistence pressure for the polarizable Stockmayer fluids 
with $|\bbox{m_0}^*|=1.0$.
The notation is the same as in figure 4.  
 }
\end{figure}

\begin{figure}
\epsfxsize=3.3in
\epsffile{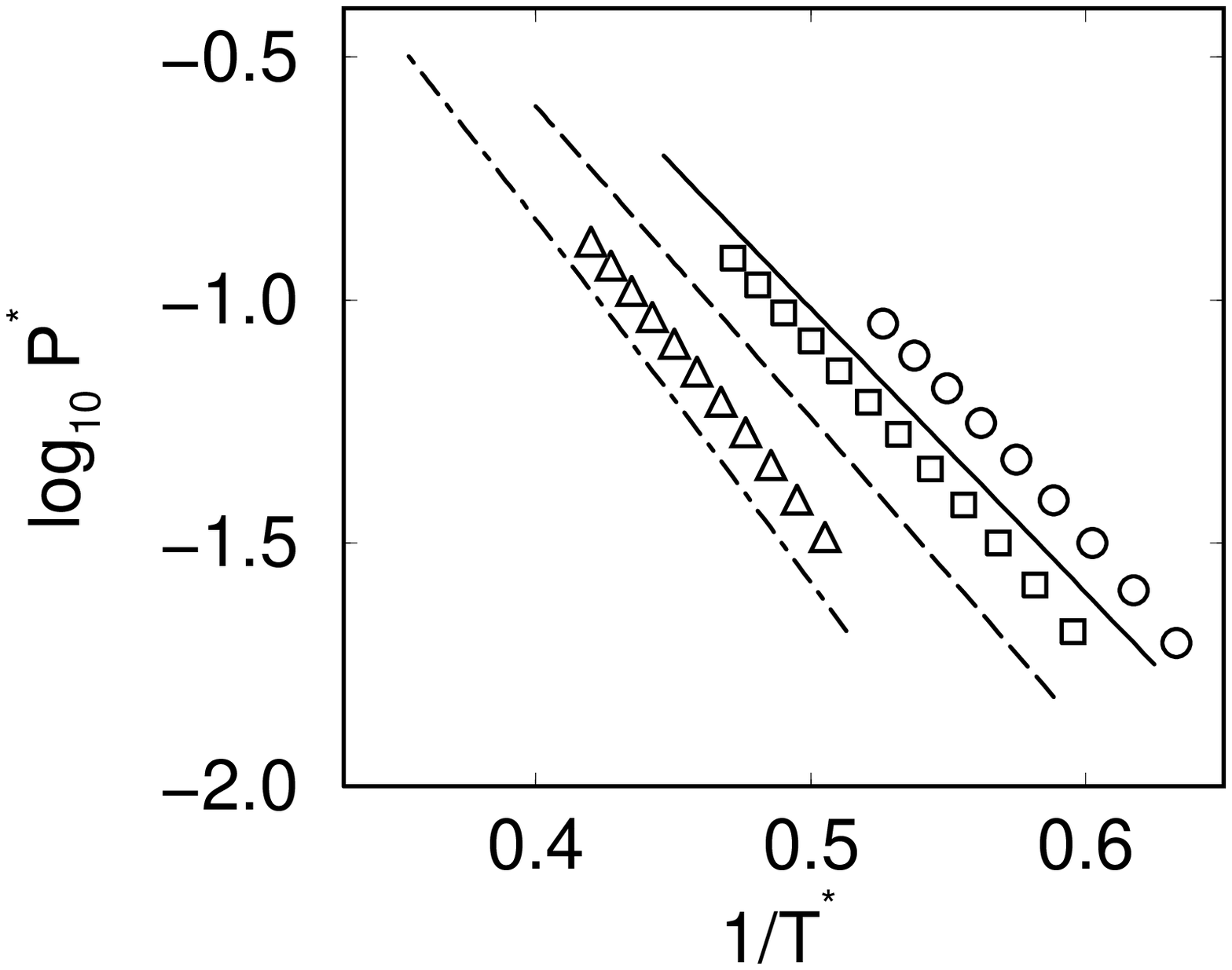}
\caption{Coexistence pressure for the polarizable Stockmayer fluids 
with $|\bbox{m_0}^*|=2.0$.  
The notation is the same as in figure 4.
}
\end{figure}

\begin{figure}
\epsfxsize=3.3in
\epsffile{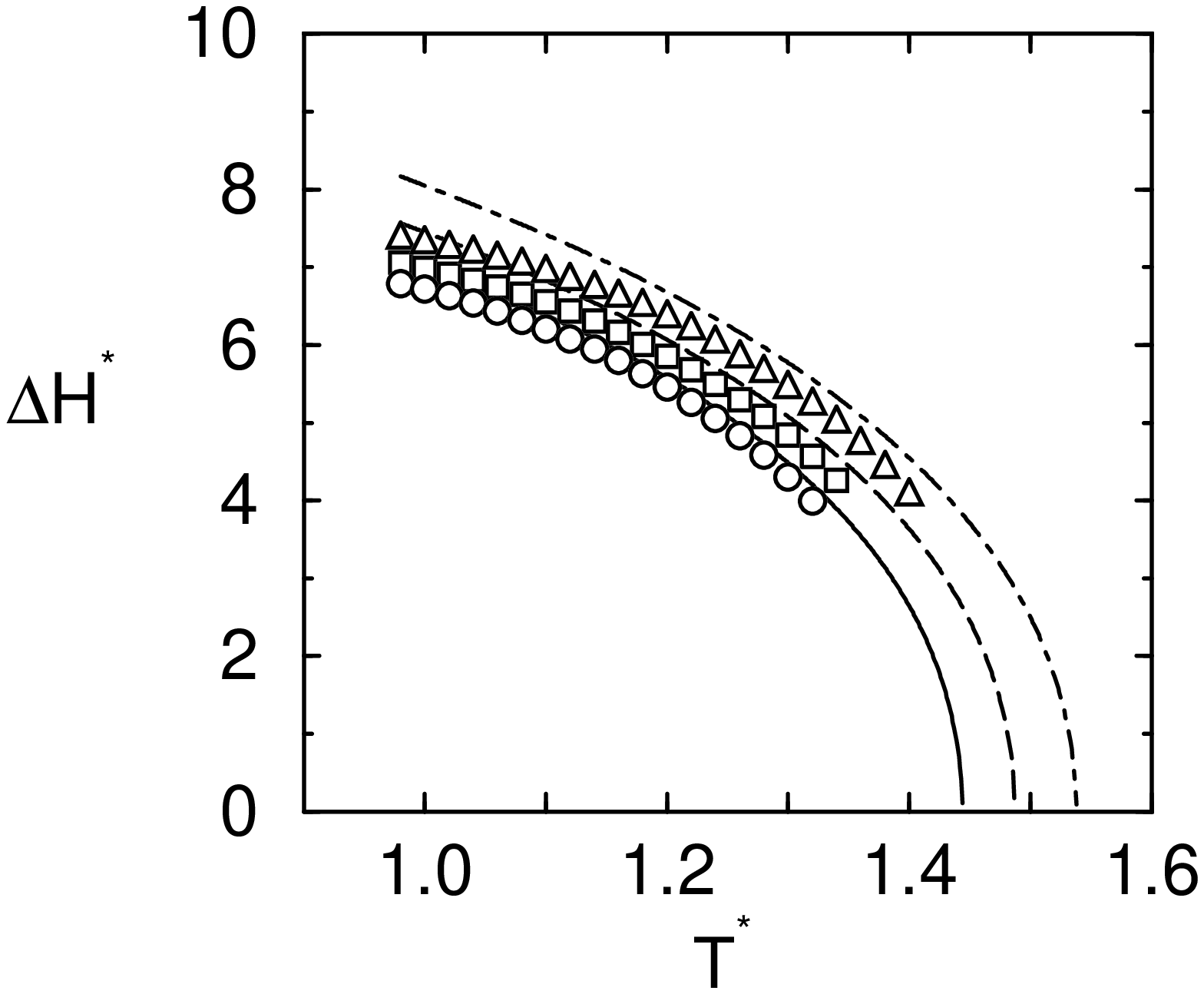}
\caption{Heat of vaporization for the polarizable Stockmayer fluids 
with $|\bbox{m_0}^*|=1.0$.
The notation is the same as in figure 4.
 }
\end{figure}

\begin{figure}
\epsfxsize=3.3in
\epsffile{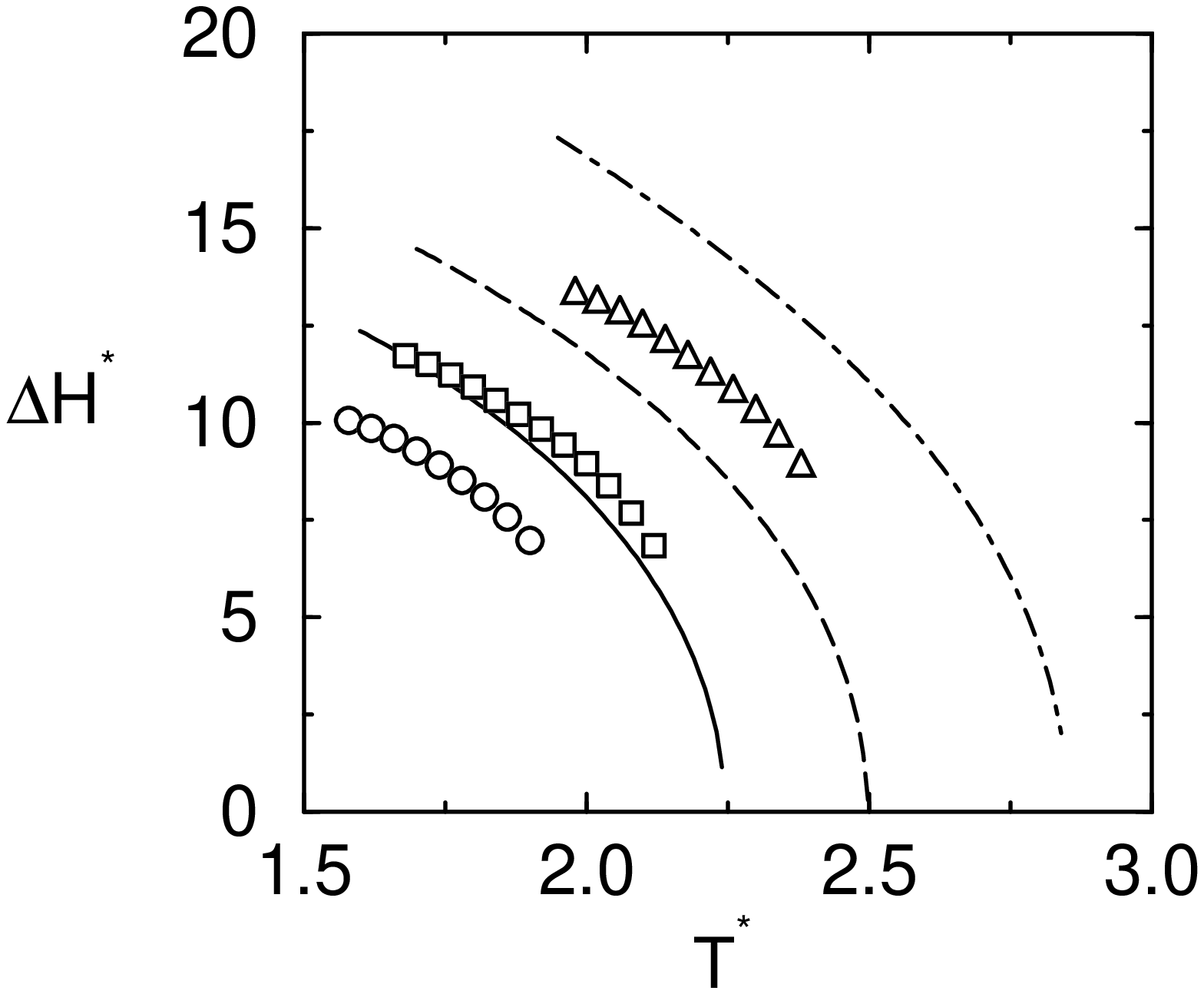}
\caption{Heat of vaporization for the polarizable Stockmayer fluids 
with $|\bbox{m_0}^*|=2.0$.  
The notation is the same as in figure 4.
 }
\end{figure}

\end{document}